\documentclass[12pt]{article}

\usepackage{sbc-template}

\usepackage{graphicx,url,tabularx,booktabs}
\usepackage[utf8]{inputenc}
\usepackage{comment}
\usepackage{ifthen}
\usepackage{graphicx}
\usepackage{tcolorbox}
\usepackage{xcolor}
\usepackage{amssymb}
\usepackage{pifont}
\usepackage[table]{xcolor}
\usepackage{textcomp}
\usepackage{listings}
\usepackage{orcidlink}
\usepackage{tikz}
\usepackage{amsmath}
\usetikzlibrary{shapes.geometric, arrows.meta, positioning, fit, backgrounds, calc}

\usepackage{pgfplots}
\pgfplotsset{compat=1.18}
\usepgfplotslibrary{colormaps}
\usetikzlibrary{matrix}
\usepackage{tikz}
\usepackage{booktabs}
\usepackage{multirow}
\usepackage{colortbl}

\newcommand{\red}[1]{\textcolor{black}{#1}}

\title{Security-First Evaluation of Text-to-Terraform: Benchmarking LLMs and SLMs for Secure IaC Generation}
\author{Francis Luis Santos Vargas\inst{1}, Rodrigo Brandão Mansilha\inst{1}\orcidlink{0000-0002-2083-653X}, Diego Kreutz\inst{1}\orcidlink{0000-0003-0830-0238}}

\address{AI Horizon Labs, PPGES, Universidade Federal do Pampa (UNIPAMPA)\\\vspace{-6mm}
\email{{\scriptsize \{francisvargas.aluno,rodrigomansilha,diegokreutz\}@unipampa.edu.br}}\\\vspace{0mm}
}

\begin{document}

\maketitle
\begin{abstract}
Cloud misconfiguration remains a leading cause of security incidents,
yet whether LLMs and SLMs can generate security-compliant
Infrastructure-as-Code is an open question. We benchmark seven models,
three closed LLMs (Claude~Opus~4, GPT-5.4, Gemini~2.5~Pro) and four
open SLMs (Qwen2.5-Coder-14B, WizardCoder-33B, CodeLlama-13B,
Magicoder-S-CL-7B), on AWS Terraform generation across 17~scenarios,
integrating Checkov and Trivy scanners into a GitLab CI/CD pipeline and
evaluating two prompt strategies at three security levels (pass@5).
Syntactic validity and security compliance are largely decoupled
properties in LLM-generated IaC, a model that reliably produces
well-formed Terraform does not necessarily produce secure Terraform:
WizardCoder-33B achieves 77.8\% validate rate yet zero Checkov
compliance, while Claude~Opus~4 reaches 23.1\% Checkov and 92.5\%
Trivy pass rates under detailed security prompting. Consequently,
prompt engineering alone is insufficient: automated multi-tool scanning
remains a necessary complement to LLM-assisted IaC generation regardless
of model family or prompt strategy.
All artifacts are publicly available.
\end{abstract}

\section{Introduction}
\label{sec:introducao}

Infrastructure-as-Code (IaC) has become the \textit{de facto} paradigm
for cloud resource provisioning, enabling reproducibility, version
control, and automated deployment~\cite{morris2020iac}. Among IaC
technologies, Terraform has emerged as the most widely adopted solution.
Yet cloud misconfiguration remains a critical and persistent threat: the
Cloud Security Alliance ranks it as the top cloud risk in
2024\footnote{\url{https://cloudsecurityalliance.org/research/topics/top-threats}}, and industry analyses attribute nearly 23\% of cloud
security incidents to incorrectly configured resources~\cite{sentinelone2024}.
Even human-authored IaC frequently introduces security smells and policy
violations~\cite{rahman2019seven,saavedra2022glitch,verdet2025emse},
and large-scale analyses of S3, VPC, and KMS deployments show that access
control and encryption misconfigurations remain major sources of data
exposure~\cite{continella2018bucket}.

LLMs and SLMs have rapidly become central to software development, with
growing adoption in code generation and infrastructure automation. Yet
AI-generated code consistently exhibits security weaknesses in
security-sensitive scenarios~\cite{pearce2022asleep,perry2023users,fang2024llms}.
Existing IaC benchmarks, IaC-Eval, DPIaC-Eval, and TerraFormer,
advance the state of the art in functional correctness and syntactic
validity, but leave a critical question open: \emph{can models that
generate valid Terraform also generate secure Terraform?} We address
this gap with four research questions:

\begin{description}

\item[\textbf{RQ1: Syntactic validity and security compliance.}]
\textit{Can state-of-the-art LLMs and locally deployable SLMs generate
syntactically valid and security-compliant AWS Terraform, and are these
two properties correlated?}
We answer RQ1 in Sections~\ref{sec:validity} and~\ref{sec:security}.

\item[\textbf{RQ2: Impact of prompt security specificity.}]
\textit{Does increasing prompt security specificity substantially improve
security compliance, and does this effect differ between LLMs and SLMs?}
We answer RQ2 in Section~\ref{sec:security}.

\item[\textbf{RQ3: Scanner-feedback self-correction.}]
\textit{Can LLMs and SLMs self-correct security-non-compliant Terraform configurations when provided with raw scanner feedback, and does correction capability differ between Checkov and Trivy findings?}
We answer RQ3 in Section~\ref{sec:r2b}.

\item[\textbf{RQ4: Longitudinal evolution.}]
\textit{Have SLM capabilities for IaC generation and security compliance
improved between IaC-Eval (2024) and this evaluation (2026)?}
We answer RQ4 in Section~\ref{sec:iac_eval}.

\end{description}

We present a security-first empirical benchmark combining automated
syntactic validation with dual-tool static security analysis (Checkov
and Trivy) in a fully reproducible GitLab CI/CD pipeline. Our main
contributions are: (i)~a reproducible benchmark with LLM-consensus
scenario selection across 17~scenarios and three AWS resource families;
(ii)~to our knowledge, the first dual-scanner (Checkov + Trivy) IaC benchmark, evaluating
seven models under two prompt strategies at three security levels with
pass@5; (iii)~empirical evidence that syntactic validity, plan executability,
and security compliance behave as three partially independent capability
dimensions in LLM-generated IaC, none of which is a reliable proxy
for the others; and
(iv)~the first longitudinal comparison of SLMs under both functional
correctness and security compliance criteria.

\section{Related Work}
\label{sec:related}

\noindent\textbf{Security of human-authored IaC.}
Prior work consistently shows that human-written IaC frequently violates
security best practices.
Rahman et al.~\cite{rahman2019seven} identified recurring security
smells in Puppet, later replicated in Ansible, Chef, and
Kubernetes~\cite{rahman2021ansible,rahman2023kubernetes}.
GLITCH~\cite{saavedra2022glitch} extended smell detection across
multiple IaC languages, while
Opdebeeck et al.~\cite{opdebeeck2023gasel} quantified smell prevalence
in Ansible repositories.
Verdet et al.~\cite{verdet2025emse} found widespread Terraform policy
non-compliance across cloud providers, and
Continella et al.~\cite{continella2018bucket} showed how S3
misconfigurations enable data exposure and resource injection.
Together, these studies establish misconfiguration as endemic in IaC and
motivate automated security analysis as a baseline requirement for
AI-generated IaC.

\noindent\textbf{Security of LLM-generated code.}
A parallel line of research investigates whether AI code generators
inherit or amplify these weaknesses.
Pearce et al.~\cite{pearce2022asleep} showed that GitHub Copilot
generates vulnerable code in 40\% of security-sensitive scenarios.
Perry et al.~\cite{perry2023users} found that developers assisted by AI
produce significantly more insecure code, while
Fang et al.~\cite{fang2024llms} identified major limitations in LLM
vulnerability reasoning.
Although these studies establish a strong expectation of security risk
in LLM-generated code, none focus on IaC, security-oriented prompting,
or LLM versus SLM comparisons.

\noindent\textbf{LLM-based IaC generation benchmarks.}
Benchmarking of LLM-generated IaC has expanded rapidly, although
security remains secondary in most studies.
IaC-Eval~\cite{kon2024iaceval} showed that Terraform generation remains
substantially harder than general code generation.
Multi-IaC-Eval~\cite{davidson2025multiiac} expanded evaluation across
multiple IaC frameworks, but without security scanning.
DPIaC-Eval~\cite{zhang2025dpiac} introduced iterative correction, yet
reported only 8.4\% security compliance after multiple feedback rounds.
Nekrasov et al.~\cite{nekrasov2025taxonomy} improved Terraform
syntactic validation using RAG, while
TerraFormer~\cite{terraformer2026} combined fine-tuning with formal
verification feedback for partial security gains.
Overall, security evaluation is usually absent or secondary to
functional correctness, complementary scanners are not combined, and
SLMs remain largely excluded from security-focused evaluation.

\noindent\textbf{Positioning this work.}
Table~\ref{tab:related} summarises prior work across eight dimensions.
The literature above reveals four gaps that motivate the present study.
First, \emph{multi-tool security coverage}: no previous IaC benchmark
integrates complementary scanners. Checkov evaluates policy compliance,
including encryption, access control, and logging, whereas Trivy
evaluates infrastructure misconfiguration through its
\texttt{AVD-AWS-*} policy catalogue, filtered in this study to
CRITICAL, HIGH, and MEDIUM severities. Our results show that these
dimensions are complementary rather than interchangeable.
Second, \emph{LLM and SLM comparison under security criteria}: prior
benchmarks either focus exclusively on LLMs or evaluate SLMs only under
functional correctness criteria. As a result, it remains unclear whether
the syntactic improvements observed in SLMs between 2024 and 2026 also
translated into security improvements.
Third, \emph{security-oriented prompt specificity}: only DPIaC-Eval and
Multi-IaC-Eval partially vary prompting strategies, but neither study
systematically isolates the effect of security-specific prompting across
both LLMs and SLMs.
Fourth, \emph{selective feedback revision}: previous iterative
correction strategies apply scanner feedback uniformly to all models.
In contrast, our approach triggers scanner-feedback revision only for
models whose compliance falls below a predefined threshold, isolating
the revision effect where security deficiencies are more pronounced.
Together, these gaps motivate a security-first benchmark in which
Checkov and Trivy compliance, rather than
\texttt{terraform validate}, define the primary evaluation criterion,
while both LLMs and SLMs are evaluated longitudinally within a fully
reproducible CI/CD pipeline.


\begin{table*}[!htp]
\centering
\caption{Positioning of this work relative to prior studies.
\checkmark~= fully addressed; $\sim$~= partially addressed; --~= not addressed or not applicable.}
\label{tab:related}
\resizebox{\textwidth}{!}{%
\begin{tabular}{lllccrcl}
\toprule
\textbf{Work} & \textbf{Venue} & \textbf{Scope} &
\textbf{LLM} & \textbf{SLM} & \textbf{\#Models} &
\textbf{Feedback loop} & \textbf{Scanner / Tool} \\
\midrule
Continella et al.~\cite{continella2018bucket}
  & ACSAC 2018     & S3 misconfiguration       & -- & -- & -- & -- & custom scanner \\
Rahman et al.~\cite{rahman2019seven}
  & ICSE 2019      & IaC smells (Puppet)       & -- & -- & -- & -- & SLIC$^\dagger$ \\
Rahman et al.~\cite{rahman2021ansible}
  & TOSEM 2021     & IaC smells (Ansible/Chef) & -- & -- & -- & -- & SLAC$^\dagger$ \\
Saavedra \& Ferreira~\cite{saavedra2022glitch}
  & ASE 2022       & IaC smells (multi-lang)   & -- & -- & -- & -- & GLITCH$^\dagger$ \\
Rahman et al.~\cite{rahman2023kubernetes}
  & TOSEM 2023     & IaC misconfig (K8s)       & -- & -- & -- & -- & empirical \\
Opdebeeck et al.~\cite{opdebeeck2023gasel}
  & MSR 2023       & IaC smells (Ansible)      & -- & -- & -- & -- & GASEL$^\dagger$ \\
Verdet et al.~\cite{verdet2025emse}
  & EMSE 2025      & IaC security (Terraform)  & -- & -- & -- & -- & Checkov + Tfsec \\
\midrule
Pearce et al.~\cite{pearce2022asleep}
  & IEEE S\&P 2022 & LLM code security         & \checkmark & -- &  1 & -- & CodeQL + manual \\
Perry et al.~\cite{perry2023users}
  & ACM CCS 2023   & LLM code security         & \checkmark & -- &  1 & -- & manual \\
Fang et al.~\cite{fang2024llms}
  & USENIX 2024    & LLM code analysis         & \checkmark & -- &  5 & -- & manual \\
\midrule
IaC-Eval~\cite{kon2024iaceval}
  & NeurIPS 2024        & IaC generation (Terraform)     & \checkmark & \checkmark & 11 & --         & OPA \\
DPIaC-Eval~\cite{zhang2025dpiac}
  & ACM FSE 2026        & IaC generation (CFN)           & \checkmark & --         &  6 & $\sim$     & Checkov \\
Multi-IaC-Eval~\cite{davidson2025multiiac}
  & arXiv 2025          & IaC generation (CFN/TF/CDK)    & \checkmark & $\sim$     &  3 & --         & CFN-Lint \\
Nekrasov et al.~\cite{nekrasov2025taxonomy}
  & ACM TOSEM 2026      & IaC generation + RAG           & \checkmark & --         &  2 & --         & -- \\
TerraFormer~\cite{terraformer2026}
  & ICSE-SEIP 2026      & IaC generation + fine-tuning   & \checkmark & \checkmark & 17 & \checkmark & Checkov \\
\midrule
\textbf{This work}
  & \textbf{---}   & \textbf{IaC security + gen.\ (Terraform)} &
  \checkmark & \checkmark & \textbf{7} & \checkmark & \textbf{Checkov + Trivy} \\
\bottomrule
\end{tabular}%
}
\begin{minipage}{\textwidth}
\smallskip
\footnotesize
$^\dagger$~SLIC, SLAC, GLITCH, and GASEL are purpose-built static analysis tools
introduced by the respective papers for smell detection in IaC scripts;
they are not general-purpose security scanners.
\end{minipage}
\end{table*}

\section{Methodology}
\label{sec:methodology}

\subsection{Benchmark Pipeline}

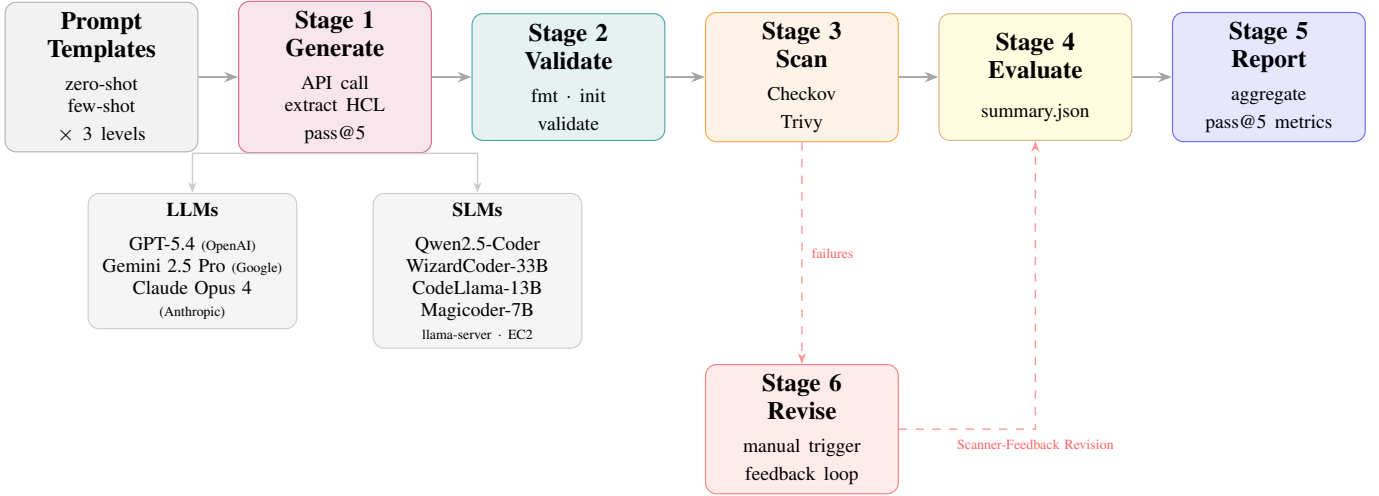
\begin{figure}[ht]
\centering
\resizebox{\columnwidth}{!}{%
\begin{tikzpicture}[
  node distance = 0.5cm,
  stage/.style = {
    rectangle,
    rounded corners=4pt,
    minimum width=2.4cm,
    minimum height=1.6cm,
    align=center,
    font=\small,
    text width=2.2cm
  },
  s1/.style={stage, draw=purple!60, fill=purple!10},
  s2/.style={stage, draw=teal!70, fill=teal!10},
  s3/.style={stage, draw=orange!70, fill=orange!10},
  s4/.style={stage, draw=brown!60, fill=yellow!15},
  s5/.style={stage, draw=blue!60, fill=blue!10},
  s6/.style={stage, draw=red!50, fill=red!8},
  inp/.style={
    rectangle,
    rounded corners=4pt,
    minimum width=2.4cm,
    minimum height=1.6cm,
    align=center,
    font=\small,
    text width=2.2cm,
    draw=gray!60,
    fill=gray!10
  },
  group/.style={
    rectangle,
    rounded corners=4pt,
    minimum width=2.6cm,
    align=center,
    font=\scriptsize,
    text width=2.4cm,
    draw=gray!40,
    fill=gray!8
  },
  arr/.style={
    -{Stealth[length=5pt,  width=10pt]},
    black,
    line width=3pt
  },
  tarr/.style={
    -{Stealth[length=5pt, width=10pt]},
    black,
    line width=2pt
  },
  darr/.style={
    -{Stealth[length=5pt, width=10pt]},
    red,
    line width=2pt,
    dashed
  }
]

\node[inp] (inp)
{
\textbf{Prompt}\\
\textbf{Templates}\\[4pt]
{\scriptsize zero-shot\\few-shot\\$\times$ 3 levels}
};

\node[s1, right=of inp] (gen)
{
\textbf{Stage 1}\\
\textbf{Generate}\\[4pt]
{\scriptsize API call\\extract HCL\\pass@5}
};

\node[s2, right=of gen] (val)
{
\textbf{Stage 2}\\
\textbf{Validate}\\[4pt]
{\scriptsize fmt $\cdot$ init\\validate}
};

\node[s3, right=of val] (scan)
{
\textbf{Stage 3}\\
\textbf{Scan}\\[4pt]
{\scriptsize Checkov\\Trivy}
};

\node[s4, right=of scan] (eval)
{
\textbf{Stage 4}\\
\textbf{Evaluate}\\[4pt]
{\scriptsize summary.json}
};

\node[s5, right=of eval] (rep)
{
\textbf{Stage 5}\\
\textbf{Report}\\[4pt]
{\scriptsize aggregate\\pass@5 metrics}
};

\node[group, below=0.5cm of gen, xshift=-1.8cm] (llms)
{
\textbf{LLMs}\\[4pt]
GPT-5.4 {\tiny (OpenAI)}\\
Gemini~2.5~Pro {\tiny (Google)}\\
Claude~Opus~4 {\tiny (Anthropic)}
};

\node[group, below=0.5cm of gen, xshift=1.8cm] (slms)
{
\textbf{SLMs}\\[4pt]
Qwen2.5-Coder\\
WizardCoder-33B\\
CodeLlama-13B\\
Magicoder-7B\\
{\tiny llama-server $\cdot$ EC2}
};

\node[s6, below=0.9cm of eval] (rev)
{
\textbf{Stage 6}\\
\textbf{Revise}\\[4pt]
{\scriptsize manual trigger\\feedback loop}
};

\draw[arr] (inp.east) -- (gen.west);
\draw[arr] (gen.east) -- (val.west);
\draw[arr] (val.east) -- (scan.west);
\draw[arr] (scan.east) -- (eval.west);
\draw[arr] (eval.east) -- (rep.west);

\draw[tarr] ([xshift=-0.5cm]gen.south) -- ++(0,-.2cm) -| (llms.north);
\draw[tarr] ([xshift=-0.5cm]gen.south)  -- ++(0,-.2cm) -|  (slms.north);

\draw[darr]
(scan.south)
|- (rev.west)
node[pos=0.25, right, font=\tiny, text=red]
{failures};

\draw[darr]
(rev.north)
-- (eval.south)
node[midway, right, font=\tiny, text=red]
{Scanner-feedback revision};

\end{tikzpicture}
}
\caption{Benchmark pipeline. Stages~1--5 run automatically as GitLab CI/CD
jobs in a parallel matrix. Stage~6 is triggered manually and applied only
to models with Round~1 pass rate below 5\% at L3.}
\label{fig:pipeline}
\end{figure}

Figure~\ref{fig:pipeline} illustrates the benchmark pipeline, implemented
as GitLab CI/CD jobs running in parallel across all evaluated models.
Stages~1--5 run automatically; Stage~6 (Revise) is triggered manually
and applied selectively to models whose Round~1 Checkov or Trivy pass
rate at L3 falls below 5\%, isolating the revision effect where
compliance gaps are most pronounced.

\subsection{Models Under Evaluation}
\label{sec:models}

Table~\ref{tab:inference_params} summarizes the model identifiers,
inference configurations, and observed truncation rates of the seven
models evaluated in this study.\footnote{Exact model files, API request
implementations, prompts, generated artefacts, and infrastructure
definitions are available in the public repository.}

Three closed-source LLMs GPT-5.4, Gemini~2.5~Pro, and
Claude~Opus~4 were accessed through REST APIs. Four open-weight SLMs
were served through \texttt{llama-server} on an AWS EC2
\texttt{c8g.16xlarge} instance equipped with AWS Graviton4 ARM64
processors, 64~vCPUs, and 128~GB of memory. All SLM inference was
performed entirely on CPU, without GPU acceleration or model-layer
offloading. The SLM subset was selected to support direct comparison
with IaC-Eval~\cite{kon2024iaceval}. Pipeline tool versions are reported
in Table~\ref{tab:tool_versions}.

\begin{table}[htb]
\centering
\caption{Pipeline tool versions.}
\label{tab:tool_versions}
\scriptsize
\setlength{\tabcolsep}{3pt}
\begin{tabular}{lll}
\toprule
\textbf{Tool} & \textbf{Version} & \textbf{Role} \\
\midrule
Terraform
  & 1.7.5
  & \texttt{fmt}, \texttt{init}, \texttt{validate}, and \texttt{plan} \\
AWS Provider
  & 6.44.0 (majority)
    & Unpinned; resolved per generation by \texttt{terraform init} \\
Checkov
  & $3.2.527$
  & \texttt{CKV\_AWS\_*} compliance scanner \\
Trivy
  & 0.69.3
  & \texttt{AVD-AWS-*} misconfiguration scanner \\
GitLab CE/EE
  & 19.0.0-pre
  & CI/CD orchestration \\
\bottomrule
\end{tabular}

\vspace{2pt}

{\scriptsize
\textit{All pipeline tools ran inside Docker containers on an AWS EC2
\texttt{c8g.16xlarge} instance using the \texttt{linux/arm64}
architecture.}
}
\end{table}

\begin{table*}[htb]
\centering
\caption{Model identifiers, inference configurations, and observed
truncation rates}\vspace{-2mm}
\label{tab:inference_params}
\scriptsize
\setlength{\tabcolsep}{2.5pt}
\begin{tabular}{lllrrrrr}
\toprule
\textbf{Model}
  & \textbf{Weights / endpoint}
  & \textbf{Quant.}
  & \textbf{Temp.}
  & \textbf{Top-p}
  & \textbf{Output cap}
  & \textbf{$n_{\mathrm{ctx}}$}
  & \textbf{Trunc.} \\
\midrule
GPT-5.4
  & \texttt{gpt-5.4}
  & --
  & \textit{default}
  & \textit{default}
  & 12{,}000
  & --
  & 0.0\% \\
Gemini~2.5~Pro
  & \texttt{gemini-2.5-pro}
  & --
  & \textit{default}
  & \textit{default}
  & 12{,}000
  & --
  & 0.0\% \\
Claude~Opus~4
  & \texttt{claude-opus-4-5}
  & --
  & \textit{default}
  & \textit{default}
  & 12{,}000
  & --
  & 0.0\% \\
\midrule
WizardCoder-33B
  & \texttt{wizardcoder-33b-v1.1-q4\_k\_m.gguf}
  & Q4\_K\_M
  & \textit{default}
  & \textit{default}
  & 12{,}000
  & 8{,}192
  & 0.0\% \\
Qwen2.5-Coder-14B
  & \texttt{qwen2.5-coder-14b-q4\_k\_m.gguf}
  & Q4\_K\_M
  & \textit{default}
  & \textit{default}
  & 12{,}000
  & 4{,}096
  & 0.0\% \\
CodeLlama-13B
  & \texttt{codellama-13b-instruct-q4\_k\_m.gguf}
  & Q4\_K\_M
  & \textit{default}
  & \textit{default}
  & 12{,}000
  & 4{,}096
  & \textbf{32.7\%} \\
Magicoder-S-CL-7B
  & \texttt{Magicoder-S-CL-7B.Q4\_K\_S.gguf}
  & Q4\_K\_S
  & \textit{default}
  & \textit{default}
  & 12{,}000
  & 4{,}096
  & 2.9\% \\
\bottomrule
\end{tabular}
 
\vspace{2pt}
 
{\scriptsize
\textit{Parameters marked as \emph{default} were not explicitly supplied
in the API requests or \texttt{llama-server} commands; therefore,
generation followed the default decoding behaviour of the corresponding
provider or inference framework. \emph{Trunc.} denotes the fraction of
generations terminated because the configured output or context limit
was reached. All SLM inference ran entirely on CPU using an AWS EC2
\texttt{c8g.16xlarge} instance equipped with AWS Graviton4 ARM64
processors. No GPU acceleration or model-layer offloading was used.
For SLMs, the effective output length was bounded by the available
context window after accounting for the input prompt.}
}
\end{table*}

\noindent\textbf{Unspecified decoding parameters.}
Temperature and top-p were not explicitly configured in either the
closed-model API requests or the \texttt{llama-server} commands.
Accordingly, the experiment followed the default decoding behavior
implemented by each provider or inference framework at evaluation time.
We report these parameters as \textit{default}, rather than assigning
retrospective values that were not recorded in the experimental
artefacts. No random seed was explicitly fixed. Instead, the
experimental protocol used $k=5$ independent generations for each
model, scenario and prompt combination to capture output variability.

\noindent\textbf{Observed truncation.}
CodeLlama-13B reached its configured 4{,}096-token context limit in
501 of 1{,}530 recorded generations (32.7\%), as indicated by
\texttt{finish\_reason=length} in the raw generation artefacts.
Magicoder-S-CL-7B reached the same limit in 45 generations (2.9\%).
The resulting outputs were forwarded to Terraform validation and
security scanning without repair, continuation, or manual
post-processing. Consequently, truncation may partly explain
CodeLlama's elevated validation-failure rate and should be considered
when comparing its results with models evaluated under larger effective
context windows.

\subsection{Scenarios and Prompt Strategies}

We evaluate 17~scenarios spanning three AWS resource families (S3, KMS,
VPC), drawn from the IaC-Eval dataset~\cite{kon2024iaceval} and filtered
via unanimous agreement (6/6) among six LLM evaluators (Claude, GPT,
Gemini~Pro, DeepSeek, Mistral, Grok). The unanimous threshold ensures that only scenarios with unambiguous 
security relevance enter the benchmark, avoiding the inclusion of 
borderline cases that could dilute the security signal.

Each scenario is evaluated under two prompt strategies at three security
levels, yielding six variants per scenario. \textbf{Zero-shot} provides
only the task description; \textbf{few-shot} adds one security-compliant
example configuration. \textbf{L1} (Without Security) serves as a
baseline measuring spontaneous security behaviour; \textbf{L2} (Basic
Security) adds general guidelines (encryption, public access blocking,
TLS); \textbf{L3} (Detailed Security) explicitly specifies Checkov
(\texttt{CKV\_AWS\_*}, \texttt{CKV2\_AWS\_*}) and Trivy
(\texttt{AVD-AWS-*}) identifiers with required resource attributes.

\subsection{Evaluation Metrics and Statistical Analysis}
\label{sec:stats}

Metrics span four dimensions. \emph{Syntactic validity}:
\texttt{terraform fmt}, \texttt{validate}, and \texttt{plan} pass rates,
with validate treated as a hard prerequisite for security analysis.
\emph{Security compliance}: Checkov pass (zero \texttt{CKV\_AWS\_*}
failures) and Trivy pass (no CRITICAL/HIGH/MEDIUM \texttt{AVD-AWS-*}
findings), computed exclusively over validate-passing generations.
\emph{Robustness}: empirical pass@5 over $k=5$ independent runs per
cell, following Chen et al.~\cite{chen2021codex} but reporting empirical
rates rather than the unbiased estimator.
\emph{Cost efficiency}: cost per run and cost per compliant generation,
using provider token pricing for LLMs.

All pass rates are accompanied by 95\% Wilson score confidence
intervals~\cite{wilson1927probable}. Pairwise model comparisons use
Pearson's chi-square on 2$\times$2 contingency tables with Bonferroni
correction ($\alpha' = 0.05/21 = 0.0024$). All reported $\chi^2$ and
$p$-values are post-correction.

\paragraph{Plan executability.}
\texttt{terraform plan} is executed over all validate-passing
generations as an additional syntactic correctness check beyond
\texttt{terraform validate}. For the three LLMs, plan was executed
as a post-processing step directly over the validated \texttt{main.tf}
artifacts produced by the main pipeline, on the same EC2 instance
with the same tooling. No additional generation was performed; plan results
reflect the same artifacts evaluated by Checkov and Trivy.
Plan\% is always computed over the validate-passing subset only.

\section{Results}
\label{sec:results}

All security metrics (Checkov, Trivy) are computed exclusively over
generations that passed \texttt{terraform validate}. Configurations
that fail validation are recorded separately and excluded from scanner
aggregations.

\subsection{Syntactic Validity and Plan Executability}
\label{sec:validity}

Table~\ref{tab:validity} reports validate and plan pass rates per model
out of 510 total generations (17~scenarios $\times$ 6~variants $\times$
5~runs).

\begin{table}[htb]
\centering
\caption{Validate and plan pass rates (510 total generations).
        }\vspace{-2mm}
\label{tab:validity}
\scriptsize
\setlength{\tabcolsep}{3pt}
\begin{tabular}{l l r r r r}
\toprule
\textbf{Model} & \textbf{Type} &
\textbf{Valid/510} & \textbf{Val\% [95\% CI]} &
\textbf{Plan/Valid} & \textbf{Plan\% [95\% CI]} \\
\midrule
GPT-5.4           & LLM & 435 & 85.3 [82.0--88.1] & 318/435 & 73.1 [68.7--77.1] \\
Claude Opus~4     & LLM & 410 & 80.4 [76.7--83.6] & 363/410 & 88.5 [85.1--91.3] \\
Gemini~2.5~Pro    & LLM & 401 & 78.6 [74.9--82.0] & 303/401 & 75.6 [71.1--79.5] \\
\midrule
WizardCoder~33B   & SLM & 397 & 77.8 [74.0--81.2] & 324/397 & 81.6 [77.5--85.1] \\
Qwen2.5-Coder~14B & SLM & 249 & 48.8 [44.5--53.2] & 205/249 & 82.3 [77.1--86.6] \\
CodeLlama~13B     & SLM & 132 & 25.9 [22.3--29.9] & 124/132 & 93.9 [88.5--96.9] \\
Magicoder~7B      & SLM & 108 & 21.2 [17.9--24.9] &  82/108 & 75.9 [67.1--83.0] \\
\bottomrule
\multicolumn{6}{l}{\textit{Plan\% over validate-passing subset. CIs: Wilson 95\%  95\% in brackets.}}
\end{tabular}
\end{table}

\textbf{Validate and plan rates are partially decoupled.}
GPT-5.4 leads in validate (85.3\%) yet records the lowest plan pass
rate among LLMs (73.1\%); Claude~Opus~4 ranks second in validate but
first in plan executability (88.5\%). Among SLMs, CodeLlama~13B
achieves the highest plan pass rate of all seven models (93.9\%
[88.5--96.9]) despite the second-lowest validate rate, 124 of its
132 validate-passing generations also pass \texttt{terraform plan}.
Qwen2.5-Coder~14B similarly achieves 82.3\% plan pass rate at only
48.8\% validate, indicating its failures concentrate in structurally
malformed outputs rather than incomplete configurations.

The dominant plan failure is \texttt{No value for required variable}
(346 of 413 total failures): models declare \texttt{variable} blocks
without \texttt{default} values, causing \texttt{terraform plan} to
abort without \texttt{-var-file}. Qwen2.5-Coder~14B exhibits this
exclusively (44/44 failures). LLMs show two additional categories:
\texttt{Error in function call} (Gemini: 19, GPT-5.4: 25, from
incorrect \texttt{jsonencode}/\texttt{toset}/\texttt{flatten} usage)
and \texttt{Self-referential block} (GPT-5.4: 4, mutually dependent
KMS+CloudWatch configurations, the same pattern seen in
validation failures, Section~\ref{sec:validate_analysis}).

\paragraph{Effect of few-shot prompting.}
Table~\ref{tab:fewshot_val} shows validate and plan pass counts by
prompt strategy. For LLMs, few-shot has negligible or slightly negative
effect on validate ($-$2 to $-$13 generations), while benefiting the
two smallest SLMs substantially (+32 for CodeLlama~13B, +20 for
Magicoder~7B). Plan executability shows a sharper pattern: few-shot
\emph{degrades} plan pass counts for all LLMs (GPT: $-$52, Gemini:
$-$51, Claude: $-$37), driven by \texttt{No value for required
variable}, the few-shot example's modular structure induces
variable declarations without defaults. The two smallest SLMs
again reverse this trend (+26 and +16 plan passes).

\begin{table}[htb]
\centering
\caption{Validate and plan pass counts: zero-shot vs.\ few-shot
         (255 generations each). Plan over validate-passing only.}
\label{tab:fewshot_val}
\scriptsize
\setlength{\tabcolsep}{3pt}
\begin{tabular}{l l r r r r r r}
\toprule
\multirow{2}{*}{\textbf{Model}} & \multirow{2}{*}{\textbf{Type}} &
\multicolumn{3}{c}{\textbf{Validate}} &
\multicolumn{3}{c}{\textbf{Plan}} \\
\cmidrule(lr){3-5}\cmidrule(lr){6-8}
& & \textbf{Zero} & \textbf{Few} & \textbf{$\Delta$} &
    \textbf{Zero} & \textbf{Few} & \textbf{$\Delta$} \\
\midrule
Claude Opus~4     & LLM & 206 & 204 & $-$2  & 200 & 163 & $-$37 \\
Gemini~2.5~Pro    & LLM & 207 & 194 & $-$13 & 177 & 126 & $-$51 \\
GPT-5.4           & LLM & 220 & 215 & $-$5  & 185 & 133 & $-$52 \\
\midrule
WizardCoder~33B   & SLM & 207 & 190 & $-$17 & 170 & 154 & $-$16 \\
Qwen2.5-Coder~14B & SLM & 130 & 119 & $-$11 & 113 &  92 & $-$21 \\
CodeLlama~13B     & SLM &  50 &  82 & \textbf{+32} &  49 &  75 & \textbf{+26} \\
Magicoder~7B      & SLM &  44 &  64 & \textbf{+20} &  33 &  49 & \textbf{+16} \\
\bottomrule
\multicolumn{8}{l}{\textit{Out of 255 per strategy. Plan $\Delta$ over validate-passing subset.}}
\end{tabular}
\end{table}

\paragraph{Answer to RQ1, Part~1 --- syntactic validity.}
LLMs produce at least one valid generation in all 102 scenario and prompt
cells (validate 78.6--85.3\%). SLMs span 21.2--77.8\%, with
CodeLlama~13B and Magicoder~7B leaving 29 and 42~cells respectively
with no scannable output. Critically, validate rate does not predict
plan executability: CodeLlama~13B ranks seventh in validate yet first
in plan pass rate (93.9\% [88.5--96.9]), while GPT-5.4 leads in
validate and trails in plan (73.1\% [68.7--77.1]).

\subsection{Effect of Prompt Security Level on Security and Compliance}
\label{sec:security}

We restrict security analysis to the \emph{shared valid subset} (Inter):
scenarios for which the model produced at least one valid generation at
all three levels, ensuring L1/L2/L3 comparisons reflect genuine
prompt-driven differences rather than scenario composition changes.
Table~\ref{tab:security} reports results with 95\% Wilson CIs.

\begin{table*}[htb]
\centering
\caption{Security compliance over the shared valid subset.
         \emph{Inter} = scenarios with at least one validate-passing
         generation at all three security levels (L1, L2, and L3);
         values are pass counts; the \emph{Inter} column and all
         reported percentages refer to L3.
         95\% Wilson CIs shown for L3. Bold = best per column group.}
\label{tab:security}
\scriptsize
\setlength{\tabcolsep}{3pt}
\begin{tabular}{l l r r r r r r r r r}
\toprule
\multirow{2}{*}{\textbf{Model}} &
\multirow{2}{*}{\textbf{Type}} &
\multirow{2}{*}{\textbf{Inter}} &
\multicolumn{4}{c}{\textbf{Checkov passes [95\% CI L3]}} &
\multicolumn{4}{c}{\textbf{Trivy passes [95\% CI L3]}} \\
\cmidrule(lr){4-7}\cmidrule(lr){8-11}
& & &
\textbf{L1} & \textbf{L2} & \textbf{L3} & \textbf{CI} &
\textbf{L1} & \textbf{L2} & \textbf{L3} & \textbf{CI} \\
\midrule
Claude Opus~4  & LLM & 134 & 0 & 4 & \textbf{31 (23.1\%)} & [16.8--31.0] & 4 & 105 & \textbf{124 (92.5\%)} & [86.8--95.9] \\
Gemini~2.5~Pro & LLM & 106 & 0 & 0 & 2 (1.9\%)  & [0.5--6.6]   & 5 &  78 &  84 (79.2\%) & [70.6--85.9] \\
GPT-5.4        & LLM & 146 & 0 & 0 & 2 (1.4\%)  & [0.4--4.9]   & 2 &  51 &  24 (16.4\%) & [11.3--23.3] \\
\midrule
WizardCoder~33B   & SLM & 121 & 0 & 0 & 0 (0.0\%) & [0.0--3.1]  & 0 & 1 & 1 (0.8\%)  & [0.1--4.5]  \\
Qwen2.5-Coder~14B & SLM &  82 & 0 & 0 & 0 (0.0\%) & [0.0--4.5]  & 0 & 2 & 6 (7.3\%)  & [3.4--15.1] \\
CodeLlama~13B     & SLM &  14 & 1 & 0 & 1 (7.1\%) & [1.3--31.5] & 0 & 1 & 1 (7.1\%)  & [1.3--31.5] \\
Magicoder~7B      & SLM &  10 & 0 & 0 & 0 (0.0\%) & [0.0--27.8] & 0 & 0 & 0 (0.0\%)  & [0.0--27.8] \\
\midrule
\multicolumn{11}{p{0.85\textwidth}}{\textit{Chi-square (Bonferroni
$\alpha'=0.0024$): Claude vs.\ GPT-5.4 Trivy L3: $\chi^2=162.4$,
$p<0.001$; Checkov L3: $\chi^2=31.8$, $p<0.001$.
All LLM--SLM comparisons at L3: $p<0.001$.
CodeLlama/Magicoder CIs wide ($n\leq14$) --- directional only.
Cramér's $V$: Claude vs.\ GPT-5.4 Trivy L3: $V=0.76$ (large);
Checkov L3: $V=0.34$ (medium). All LLM--SLM pairs at L3:
$V>0.45$.}} \\
\end{tabular}
\end{table*}

\textbf{LLM compliance responds strongly to prompt specificity.}
Claude~Opus~4 shows the clearest progression: Checkov
0$\to$4$\to$31 and Trivy 4$\to$105$\to$124, the strongest security
profile in the benchmark. Gemini~2.5~Pro reaches 79.2\% Trivy at L3
despite only 1.9\% Checkov, while its per-level Inter population shrinks from 150
at L1 to 106 at L3, prompt complexity simultaneously reduces
syntactic output and improves security. GPT-5.4 peaks at Trivy L2
(51 passes) and \emph{declines} at L3 (24/146): detailed security prompts
introduce resource constructs (KMS-encrypted CloudWatch log groups)
that generate dependency graphs Trivy penalises due to resolution
failures. All LLM--SLM differences at L3 are statistically significant
($p<0.001$, Bonferroni-corrected); the Claude vs.\ GPT-5.4 Trivy
difference alone yields $\chi^2=162.4$. A per-scenario breakdown of Trivy compliance at L3 is provided
in Table~\ref{tab:heatmap} (Section~\ref{sec:heatmap}).

\textbf{SLM compliance is near-zero and unresponsive to prompting.}
All four SLMs record near-zero Checkov compliance at all levels.
The sole exception is CodeLlama~13B with 1~Checkov pass at L1 and
L3, the only non-zero SLM Checkov result in the benchmark,
but its Inter of $n=14$ renders this directional only [1.3--31.5].
WizardCoder~33B achieves 77.8\% validate yet zero Checkov passes
[0.0--3.1], the clearest evidence that syntactic correctness and
security compliance are decoupled properties. Few-shot prompting has
negligible effect on security compliance once the security level is
held constant ($\pm$1--4 passes across all models and strategies).

\paragraph{Checkov--Trivy discordance.}
Table~\ref{tab:discordance} classifies each L3 generation into four
compliance states. Unlike Table~\ref{tab:security}, this analysis is
computed over \emph{all} validate-passing L3 generations, without the
Inter restriction: conditioning a scanner-agreement analysis on
scenarios that validated at all three security levels would bias it
toward configurations the model already finds easy, which is the
wrong conditioning for measuring discordance. This explains why the
SLM denominators here (122, 94, 23, 21) exceed the corresponding
Inter values in Table~\ref{tab:security} (121, 82, 14, 10).
\emph{No model ever passes Checkov without also
passing Trivy}, checkov-only is zero across all models, confirming
Checkov is a strictly harder criterion. The dominant LLM state at L3
is Trivy-only: Claude~Opus~4 places 69.4\% of its L3 generations
here, Gemini 77.4\%. These configurations pass all CRITICAL/HIGH/MEDIUM
Trivy checks but fail at least one Checkov policy, and may
represent an acceptable posture in non-production environments.
GPT-5.4 diverges sharply: 83.6\% fail both tools vs.\ 7.5\% for
Claude and 20.8\% for Gemini, consistent with its higher rate of
architecturally complex failures. SLMs cluster uniformly in
``neither'': 91--100\% across all four models. \red{The observed discordance also highlights that infrastructure security
cannot be adequately characterised by a single analysis tool.
Configurations considered acceptable by vulnerability-oriented scanning
may still violate organisational security policies enforced by
compliance scanners.} \red{This finding also explains why previous IaC benchmarks based on a single scanner or validation criterion may overestimate the practical
security capability of current LLMs.}

\begin{table}[htb]
\centering
\caption{Checkov--Trivy compliance states at L3 (validate-passing
         generations).}
\label{tab:discordance}
\scriptsize
\setlength{\tabcolsep}{3pt}
\begin{tabular}{l l r r r r r}
\toprule
\textbf{Model} & \textbf{Type} & \textbf{$n$} &
\textbf{Both} & \textbf{Trv only} & \textbf{Chk only} & \textbf{Neither} \\
\midrule
Claude Opus~4     & LLM & 134 & 31 (23.1\%) & 93 (69.4\%) & 0 & 10 (7.5\%)   \\
Gemini~2.5~Pro    & LLM & 106 &  2 (1.9\%)  & 82 (77.4\%) & 0 & 22 (20.8\%)  \\
GPT-5.4           & LLM & 146 &  2 (1.4\%)  & 22 (15.1\%) & 0 & 122 (83.6\%) \\
\midrule
WizardCoder~33B   & SLM & 122 & 0 &  1 (0.8\%) & 0 & 121 (99.2\%)  \\
Qwen2.5-Coder~14B & SLM &  94 & 0 &  6 (6.4\%) & 0 &  88 (93.6\%)  \\
CodeLlama~13B     & SLM &  23 & 1 (4.3\%) & 1 (4.3\%) & 0 & 21 (91.3\%) \\
Magicoder~7B      & SLM &  21 & 0 & 0 & 0 & 21 (100\%) \\
\bottomrule
\end{tabular}
\end{table}

\paragraph{Answer to RQ1, Part~2 --- security compliance.}
Syntactic validity and security compliance are largely \emph{decoupled}. \red{This decoupling suggests that current code-oriented foundation models
primarily learn Terraform syntax and resource composition, while
security best practices remain only partially represented in their
internal knowledge. Consequently, generating syntactically correct IaC
appears substantially easier than generating policy-compliant IaC,
reinforcing that deployment readiness cannot be inferred from
successful Terraform validation alone.}:
WizardCoder~33B achieves 77.8\% validate yet 0.0\% Checkov [0.0--3.1].
All LLM--SLM differences at L3 are statistically significant
($p<0.001$, Bonferroni-corrected).

\paragraph{Answer to RQ2.}
Prompt specificity substantially improves LLM compliance (Claude:
$+$23.1pp Checkov, $+$89.6pp Trivy from L1$\to$L3; Gemini: $+$76.0pp
Trivy) but is ineffective for SLMs, where L3 prompts frequently degrade
syntactic validity without compliance gains. The binding constraint for
SLMs is instruction-following capacity, not prompt content.

\subsection{Scanner-Feedback Revision}
\label{sec:r2b}

After Round~1, models received their own raw Checkov and Trivy output
as corrective feedback and regenerated the configuration.
Stage~6 was applied to all Round~1 generations that produced
at least one Checkov or Trivy finding; generations that passed
both scanners in Round~1 were not revised, as no corrective
feedback was available. This accounts for the smaller revision
$n$ observed for Claude~Opus~4 and Gemini~2.5~Pro relative
to their Round~1 $n$.
Table~\ref{tab:r2b} compares normalised pass rates before and after
this single-pass revision.

\begin{table}[htb]
\centering
\caption{Round~1 vs.\ scanner-feedback revision.
         Chk\% and Trv\% = passes / $n$ validate-passing generations.}
\label{tab:r2b}
\scriptsize
\setlength{\tabcolsep}{3pt}
\begin{tabular}{l l r r r r r r r}
\toprule
\textbf{Model} & \textbf{Type} &
\multicolumn{3}{c}{\textbf{Round~1}} &
\multicolumn{4}{c}{\textbf{Revision}} \\
\cmidrule(lr){3-5}\cmidrule(lr){6-9}
& & \textbf{$n$} & \textbf{Chk\%} & \textbf{Trv\%} &
    \textbf{$n$} & \textbf{Chk} & \textbf{Chk\%} & \textbf{Trv\%} \\
\midrule
Claude Opus~4     & LLM & 410 &  8.5\% & 56.8\% & 346 &  72 & \textbf{20.8\%} & 49.7\% \\
Gemini~2.5~Pro    & LLM & 401 &  0.5\% & 42.9\% & 341 &  34 & 10.0\% & 39.0\% \\
GPT-5.4           & LLM & 435 &  0.5\% & 17.7\% & 355 &  10 &  2.8\% & \textbf{61.4\%} \\
\midrule
WizardCoder~33B   & SLM & 397 &  0.0\% &  0.5\% & 171 &   0 &  0.0\% & 42.1\% \\
Qwen2.5-Coder~14B & SLM & 249 &  0.0\% &  3.2\% &  53 &   0 &  0.0\% &  5.7\% \\
CodeLlama~13B     & SLM & 132 &  3.0\% &  3.8\% &  52 &   2 &  3.8\% & 26.9\% \\
Magicoder~7B      & SLM & 108 &  0.0\% &  0.9\% &  38 &   1 &  2.6\% & 13.2\% \\
\bottomrule
\multicolumn{9}{l}{\textit{Trivy decline for Claude/Gemini: subset composition differs between rounds.}}
\end{tabular}
\end{table}

Trivy compliance improves substantially across most models after
feedback: GPT-5.4 gains $+$43.7pp (17.7\%$\to$61.4\%),
WizardCoder~33B $+$41.6pp (0.5\%$\to$42.1\%), CodeLlama~13B
$+$23.1pp. This confirms that Trivy findings, concrete severity
labels and resource paths, provide actionable correction signals
that models can directly incorporate. Checkov compliance improves
only for LLMs: Claude~Opus~4 gains $+$12.3pp
(8.5\%$\to$20.8\%), Gemini $+$9.5pp. Checkov violations typically
require architectural decisions (adding encryption resources,
configuring IAM, enabling logging) that a single-pass revision
cannot fully resolve. The Trivy decline for Claude and Gemini reflects a subset
composition artefact: the revision subset ($n=346$ and $n=341$)
excludes the 64 and 60 Round~1 generations that already passed
both scanners and therefore received no corrective feedback.
Since these clean-passing generations are removed from the
revision denominator, the revision Trivy rate is computed over
a strictly harder subset than Round~1, producing an apparent
decline that does not reflect genuine degradation.

\paragraph{Answer to RQ3.}
Models partially self-correct from raw scanner output, but capability
differs systematically by tool and model family. Trivy findings are
broadly actionable across both families; Checkov gains are confined to
the LLMs, with GPT-5.4 and all SLMs showing minimal improvement. The
asymmetry reflects the structural difference between localised,
resource-scoped findings and systemic compliance gaps: models act on
the former but not reliably on the latter.

\subsection{Longitudinal Comparison with IaC-Eval}
\label{sec:iac_eval}

Three SLMs were previously evaluated in IaC-Eval~\cite{kon2024iaceval}
(NeurIPS 2024) under a functional correctness criterion assessed via
\texttt{terraform plan} and OPA Rego policies. Our evaluation reuses
17 of the 458 IaC-Eval scenarios and extends assessment with Checkov
and Trivy. The IaC-Eval dataset encodes a \emph{generation difficulty}
label per scenario, but this axis does not translate to security:
a scenario easy to implement correctly may be consistently generated
insecurely, as confirmed by WizardCoder~33B's 77.8\% validate rate
at zero Checkov compliance. Table~\ref{tab:iac_eval} places both
benchmarks side by side.

\begin{table}[htb]
\centering
\caption{Longitudinal comparison with IaC-Eval~\cite{kon2024iaceval}
         (458 scenarios, pass@1, functional correctness).
         This work: 17 scenarios, pass@5, syntactic validity $+$
         plan executability $+$ security.
         $\dagger$ = model also in IaC-Eval.}\vspace{-2mm}
\label{tab:iac_eval}
\scriptsize
\setlength{\tabcolsep}{3pt}
\begin{tabular}{l l r r r r r}
\toprule
\textbf{Model} & \textbf{Type} &
\textbf{IaC-Eval} & \textbf{Val\%} \textbf{pass@1} & \textbf{Plan\%} &
\textbf{Chk/$n$ (L3)} & \textbf{Trv/$n$ (L3)} \\
\midrule
Claude Opus~4     & LLM & ---    & 80.4\% & 88.5\% & 31/134 & 124/134 \\
Gemini~2.5~Pro    & LLM & ---    & 78.6\% & 75.6\% &  2/106 &  84/106 \\
GPT-5.4           & LLM & ---    & 85.3\% & 73.1\% &  2/146 &  24/146 \\
\midrule
WizardCoder~33B$^\dagger$   & SLM & 8.93\% & 77.8\% & 81.6\% & 0/121 &  1/121 \\
Qwen2.5-Coder~14B & SLM & ---    & 48.8\% & 82.3\% & 0/ 82 &  6/ 82  \\
CodeLlama~13B$^\dagger$     & SLM & 2.01\% & 25.9\% & 93.9\% & 1/ 14 &  1/ 14  \\
Magicoder~7B$^\dagger$      & SLM & 7.62\% & 21.2\% & 75.9\% & 0/ 10 &  0/ 10  \\
\bottomrule
\multicolumn{7}{l}{\textit{Plan\% for LLMs: results\_plan\_llms.json (this work).}}
\end{tabular}
\end{table}

\textbf{Syntactic improvement does not imply security improvement.}
Validate rates are substantially higher than IaC-Eval pass@1 for the
same SLMs (WizardCoder: 77.8\% vs.\ 8.93\%; Magicoder: 21.2\%
vs.\ 7.62\%), as expected: IaC-Eval requires functional correctness
in addition to compilation, a strictly harder criterion applied over
the full 458-scenario dataset vs.\ our 17-scenario subset.

\textbf{Plan executability inverts the validate ranking.}
CodeLlama~13B achieves 93.9\% plan pass rate despite the second-lowest
validate rate, while GPT-5.4 leads in validate yet has the lowest plan
pass rate among LLMs (73.1\%). This inversion confirms that validate
rate, plan executability, and security compliance are three distinct
and partially independent properties, none is a reliable proxy for
the others.

\textbf{Security capability has not improved alongside generation quality.}
None of the SLMs previously evaluated in IaC-Eval achieve meaningful
security compliance: WizardCoder~33B, the highest-ranked SLM in
IaC-Eval (8.93\%), records 0 Checkov and only 1 Trivy pass out of
121 valid L3 generations. Magicoder~7B and CodeLlama~13B show
equally marginal results. The IaC generation capability improvements
observed between 2023 and 2026 have not translated into security-aware
generation for SLMs.

\paragraph{Answer to RQ4.}
Between 2024 and 2026, SLMs became better at generating Terraform;
they did not become better at generating \emph{secure} Terraform.
Syntactic generation improved substantially, but security compliance
remained near-zero and unresponsive to prompt specificity, the
binding constraint is model architecture and training, not prompting
strategy. \red {These findings suggest that recent advances in code-generation models
have predominantly improved Terraform syntax generation rather than
security-aware infrastructure design. Future improvements in secure IaC
generation will likely depend more on security-specific training,
policy-guided optimisation, or retrieval mechanisms than on increasing
general model capability alone.}

\subsection{Per-Scenario Compliance Pattern}
\label{sec:heatmap}

Table~\ref{tab:heatmap} reports Trivy compliance at L3 (Detailed
Security) per model and scenario pair. Each cell shows \emph{pass/valid},
where \emph{pass} is the number of validate-passing generations that
also passed Trivy, and \emph{valid} is the total validate-passing
generations at L3 across both prompt strategies and 5~runs (maximum
10 per cell). Cells where \emph{valid} $<$ 10 reflect validate
failures at L3 for that model and scenario combination; cells marked
``---'' indicate zero validate-passing generations.

\newcommand{\hh}[2]{%
  \ifnum #2=0 \cellcolor{gray!12}{\tiny ---}%
  \else\ifnum #1=0 \cellcolor{red!8}{\tiny 0/#2}%
  \else
    \pgfmathparse{#1/#2*100}%
    \ifdim \pgfmathresult pt > 74pt
      \cellcolor{green!40}{\tiny\textbf{#1/#2}}%
    \else\ifdim \pgfmathresult pt > 49pt
      \cellcolor{green!25}{\tiny #1/#2}%
    \else
      \cellcolor{orange!30}{\tiny #1/#2}%
    \fi\fi
  \fi\fi}

\begin{table*}[htb]
\centering
\caption{Trivy pass/valid at L3 per model--scenario.
         \emph{pass} = validate-passing generations that passed Trivy;
         \emph{valid} = total validate-passing generations at L3
         (max 10 = 2~strategies $\times$ 5~runs; lower values
         reflect validate failures for that model--scenario).
         S1--S9: S3/KMS; S10--S17: VPC.
         \colorbox{green!40}{\textbf{$\geq$75\%}} /
         \colorbox{green!25}{50--74\%} /
         \colorbox{orange!30}{1--49\%} /
         \colorbox{red!8}{0\%} /
         \colorbox{gray!12}{---} = no validate-passing generations at L3.}
\label{tab:heatmap}
\resizebox{\textwidth}{!}{%
\scriptsize
\setlength{\tabcolsep}{1.5pt}
\begin{tabular}{l l *{17}{>{\centering\arraybackslash}p{0.55cm}}}
\toprule
\textbf{Model} & \textbf{T} &
\textbf{S1} & \textbf{S2} & \textbf{S3} & \textbf{S4} &
\textbf{S5} & \textbf{S6} & \textbf{S7} & \textbf{S8} &
\textbf{S9} & \textbf{S10} & \textbf{S11} & \textbf{S12} &
\textbf{S13} & \textbf{S14} & \textbf{S15} & \textbf{S16} & \textbf{S17} \\
\midrule
Claude    & L & \hh{8}{8} & \hh{8}{8} & \hh{8}{8} & \hh{8}{8} & \hh{8}{8} & \hh{7}{7} & \hh{7}{8} & \hh{8}{8} & \hh{8}{8} & \hh{5}{7} & \hh{8}{8} & \hh{8}{8} & \hh{8}{8} & \hh{8}{8} & \hh{8}{8} & \hh{6}{8} & \hh{3}{8} \\
Gemini    & L & \hh{4}{7} & \hh{10}{10} & \hh{8}{8} & \hh{8}{9} & \hh{8}{10} & \hh{9}{9} & \hh{5}{6} & \hh{7}{7} & \hh{4}{4} & \hh{0}{3} & \hh{2}{3} & \hh{2}{4} & \hh{4}{7} & \hh{1}{2} & \hh{9}{10} & \hh{2}{2} & \hh{1}{5} \\
GPT-5.4   & L & \hh{0}{8} & \hh{7}{8} & \hh{7}{8} & \hh{6}{8} & \hh{0}{10} & \hh{1}{8} & \hh{0}{10} & \hh{0}{10} & \hh{0}{10} & \hh{0}{8} & \hh{0}{9} & \hh{0}{10} & \hh{0}{7} & \hh{0}{8} & \hh{0}{8} & \hh{2}{8} & \hh{0}{8} \\
\midrule
WizardCdr & S & \hh{0}{8} & \hh{0}{9} & \hh{1}{7} & \hh{0}{8} & \hh{0}{5} & \hh{0}{9} & \hh{0}{1} & \hh{0}{9} & \hh{0}{1} & \hh{0}{7} & \hh{0}{8} & \hh{0}{10} & \hh{0}{9} & \hh{0}{9} & \hh{0}{7} & \hh{0}{7} & \hh{0}{8} \\
Qwen      & S & \hh{0}{5} & \hh{0}{5} & \hh{3}{6} & \hh{0}{4} & \hh{2}{9} & \hh{1}{6} & \hh{0}{1} & \hh{0}{6} & \hh{0}{4} & \hh{0}{3} & \hh{0}{0} & \hh{0}{10} & \hh{0}{9} & \hh{0}{10} & \hh{0}{0} & \hh{0}{8} & \hh{0}{8} \\
CodeLlama & S & \hh{0}{0} & \hh{0}{1} & \hh{0}{3} & \hh{0}{1} & \hh{0}{3} & \hh{0}{1} & \hh{0}{0} & \hh{1}{1} & \hh{1}{1} & \hh{0}{2} & \hh{0}{4} & \hh{0}{1} & \hh{0}{3} & \hh{0}{0} & \hh{0}{1} & \hh{0}{0} & \hh{0}{1} \\
Magicoder & S & \hh{0}{0} & \hh{0}{1} & \hh{0}{0} & \hh{0}{3} & \hh{0}{3} & \hh{0}{3} & \hh{0}{1} & \hh{0}{1} & \hh{0}{0} & \hh{0}{0} & \hh{0}{1} & \hh{0}{1} & \hh{0}{1} & \hh{0}{2} & \hh{0}{1} & \hh{0}{1} & \hh{0}{2} \\
\bottomrule
\end{tabular}%
}
\par\vspace{2pt}
{\tiny T = model type (L = LLM, S = SLM).
Scanning applied only to validate-passing generations;
validate failures reduce the denominator below 10.}
\end{table*}

Two structural patterns emerge. First, \emph{compliance difficulty
is scenario-driven, not only model-driven}: S17 (VPC with two subnets
and routing tables) records 3/8 for Claude and near-zero for all
other models despite all having validate-passing generations;
S10 (VPC subnet association) drops Claude to 5/7. Both scenarios
require VPC flow log configuration and default security group
restrictions that models consistently omit regardless of prompt
level. Second, \emph{GPT-5.4's Trivy compliance is structurally
concentrated in simple S3 scenarios}: cells S2--S4 show 6--8/8,
while all KMS and VPC scenarios record 0 or near-zero, revealing
that its aggregate compliance rate masks near-complete failure on
architecturally complex configurations, consistent with the
self-referential block errors identified in
Section~\ref{sec:validate_analysis}.

\subsection{Analysis of Validation Failures}
\label{sec:validate_analysis}

We identify three failure categories from \texttt{terraform validate}
error messages across all models, strategies, and levels.

\noindent\textbf{Category~1 --- Semantic dependency errors
(Claude~Opus~4, GPT-5.4).}
Both models predominantly fail with \textit{Reference to undeclared
resource} in complex VPC+KMS scenarios: they generate cross-resource
references correct in intent but absent from the configuration.
GPT-5.4 additionally produces \textit{Resource cycle} errors
(\texttt{aws\_kms\_key $\leftrightarrow$ aws\_cloudwatch\_log\_group}), a semantically correct security pattern that Terraform cannot
instantiate without an intermediate \texttt{data} source. These
failures reveal an \emph{ambition gap}: both models attempt
architecturally sophisticated, security-correct configurations but
produce dependency graphs that require additional scaffolding to
resolve.

\noindent\textbf{Category~2 --- Provider API version mismatches
(Gemini~2.5~Pro, WizardCoder~33B, Qwen2.5-Coder~14B).}
These models generate deprecated AWS provider syntax:
\texttt{versioning\_configuration} as a nested block inside
\texttt{aws\_s3\_bucket} (removed in provider v4+), deprecated
\texttt{acl} attributes, and \texttt{required\_providers} blocks
without \texttt{source}/\texttt{version} fields. Qwen2.5-Coder~14B
additionally generates module-style \texttt{var.*} references in
non-module contexts. The pattern reflects training data staleness
relative to the provider major version resolved at run time, which
was 6.x for most generations.

\noindent\textbf{Category~3 --- Structural HCL errors
(CodeLlama~13B, Magicoder~7B).}
These models produce high volumes of parser-level failures:
\textit{Invalid multi-line string}, \textit{Argument or block
definition required}, \textit{Unsupported block type}. Both also
emit \texttt{var.*} references without \texttt{variable} declarations
and occasionally leak markdown fences into the HCL output.
CodeLlama~13B additionally targets Terraform~0.11/0.12 syntax in
some generations. These failures occur before any provider-specific
validation, reflecting fundamental HCL generation deficiencies rather
than knowledge staleness.

Failures in complex scenarios (KMS encryption, VPC flow logs with
CloudWatch) leave the security-most-relevant cells unscored, creating
a systematic blind spot: models that fail precisely where security
matters most are neither penalised nor credited. The Category~1
pattern further suggests that LLM failures here reflect architectural
Terraform knowledge gaps rather than security unawareness.

\subsection{Threats to Validity}
\label{sec:threats}

\noindent\textbf{Statistical power.}
The 3{,}570 evaluations ($7 \times 17 \times 6 \times 5$) support
aggregate inference, with all pass rates accompanied by Wilson 95\%
CIs and pairwise comparisons Bonferroni-corrected. However,
CodeLlama~13B and Magicoder~7B have Inter subsets of $n=14$ and
$n=10$, yielding wide intervals ([1.3--31.5] and [0.0--27.8])
that support only directional interpretation. The $k=5$ design is
insufficient for scenario-level analysis; future work should increase
$k$ for low-compliance models.
The claim that syntactic validity, plan executability, and security
compliance form partially independent dimensions (contribution~iii)
is based on empirical observation across seven models: models that
rank highly on validate rate do not consistently rank highly on
plan pass rate or Checkov compliance, and vice versa
(Tables~\ref{tab:validity} and~\ref{tab:security}). Formal
independence testing via rank correlation (e.g., Spearman's $\rho$
across models) is left for future work with a larger model sample,
as $n=7$ models is insufficient for reliable rank correlation
estimates.

\noindent\textbf{Scanner scope and complementarity.}
Checkov targets AWS policy compliance (encryption, access controls,
logging) while Trivy applies its own misconfiguration catalogue,
filtered here to CRITICAL/HIGH/MEDIUM. Our discordance analysis
(Table~\ref{tab:discordance}) shows these are not interchangeable:
Checkov implies Trivy but not vice versa, and the Trivy-only category
(dominant for LLMs at L3) represents configurations that may be
acceptable in non-production environments. Both scanners perform
static analysis only and cannot detect runtime-dependent
vulnerabilities.

\noindent\textbf{Scenario selection and circularity.}
The 17~scenarios were selected via unanimous LLM consensus, which
may over-represent patterns in model training corpora. To partially
validate representativeness, the dominant failed Checkov checks at L3
(counted as finding occurrences),
\texttt{CKV2\_AWS\_62} (S3 event notifications, 583 failures),
\texttt{CKV\_AWS\_144} (S3 cross-region replication, 534),
\texttt{CKV2\_AWS\_11} (VPC flow logging, 229), correspond to
checks consistently identified as high-prevalence in real-world IaC
repositories~\cite{verdet2025emse}, providing partial external
validation of scenario relevance.

\noindent\textbf{Model versions, quantization, and non-determinism.}
Results reflect seven specific model versions at a single point
(May~2026) and may not extend to future releases or fine-tuned
variants. SLMs run at 4-bit quantization (Q4\_K\_M; Q4\_K\_S for Magicoder) (estimated 1--3pp
penalty vs.\ full precision~\cite{frantar2022gptq}), substantially
below the observed LLM--SLM compliance gap, suggesting architecture
and training are the primary drivers. All models were queried with the
decoding settings recorded in Table~\ref{tab:inference_params}, leaving
temperature and top-p at provider or \texttt{llama-server} defaults;
the observed pass rates therefore reflect a combination of model
architecture, training stochasticity, and provider-specific sampling
defaults, and should not be interpreted as pure measures of
architectural consistency. The IaC-Eval generation difficulty label
does not predict security compliance: scenarios easy to implement
correctly are not necessarily easy to implement securely.

\noindent\textbf{Provider resolution.}
The AWS provider was not pinned: each generation declared its own
constraint, and \texttt{terraform init} resolved it against the
registry at run time. Among preserved init records, 82\% resolved to
6.44.0 and 14\% to 5.100.0, with a residue on 3.x/4.x. Validate
outcomes are therefore not attributable to a single provider version,
and a pinned lock file should be preserved in future runs.

\section{Conclusions and Future Work}
\label{sec:conclusion}

This work presented a reproducible security-first benchmark for
evaluating the ability of large and small language models to generate
secure Terraform configurations. Through four complementary research
questions, we investigated the relationship between syntactic
correctness, security compliance, prompt sensitivity, iterative
repair, and the evolution of AI-generated Infrastructure as Code.
Overall, our findings show that syntactic validity alone is not a
reliable indicator of secure IaC generation, and that security
compliance should be treated as a first-class evaluation criterion
rather than a by-product of successful deployment. These results
converge on a single practical implication: automated multi-tool
static analysis is not an optional complement to AI-assisted IaC
generation, but a prerequisite for deployment-ready infrastructure,
regardless of model family or prompting strategy. While prompt
engineering and scanner feedback can improve generation quality, they
are insufficient to consistently eliminate security compliance gaps.
By making the benchmark, pipeline, prompts, and evaluation artefacts
publicly available, we hope this work provides a reproducible
foundation for future research on secure AI-assisted Infrastructure as
Code and supports both researchers and practitioners in developing and
evaluating more secure generation approaches.

Several research directions naturally emerge from this benchmark. From
the benchmark perspective, future versions should expand the coverage
to additional IaC languages, cloud providers, AWS resource families,
and official Terraform modules as reference implementations. From the
pipeline perspective, we plan to evolve the evaluation framework
through an MCP Server capable of continuously assessing newly released
foundation models and security tools, enabling long-term and
reproducible benchmarking. At the model level, domain-specific
fine-tuning strategies for SLMs, retrieval-augmented generation, and
policy-aware guidance represent promising directions to reduce the gap
between syntactic correctness and security compliance identified in
this study. We hope these efforts contribute to establishing
reproducible evaluation practices and accelerate the adoption of
AI-assisted Infrastructure as Code without compromising infrastructure
security.

All pipeline code, prompts, scenario definitions, and evaluation
artefacts are publicly
available.\footnote{{\tiny \url{https://gitlab.com/security-iac/security-first-evaluation-of-text-to-terraform/-/tags/sbseg-trilha-principal}}}

\section*{AI Usage Declaration}

Generative AI tools were used as support during the development of this
work. The Claude assistant (Anthropic) was employed to assist with the
revision and refinement of academic writing in English, the structuring
of paper sections, and the generation of \LaTeX{} code snippets. All
scientific content, including the experimental design, result analysis,
and conclusions, was elaborated and verified by the authors. The authors
assume full responsibility for the published content.

\section*{Acknowledgements}
This research was partially supported by Capes, CNPq (Grant No. 409743/2025-9), Fundação de Amparo à Pesquisa do Estado do Rio Grande do Sul (FAPERGS) – grants no. 24/2551-0001368-7, 24/2551-0000726-1, and 25/2551-0002572-9. It was also financed in part by Fundação de Amparo à Pesquisa do Estado de São Paulo (FAPESP)  –  grants no. \#2020/05183-0 and \#2023/00816-2. This research was partially supported by Amazon Web Services (AWS) through AWS Promotional Credits, which provided the cloud computing infrastructure used for the experimental evaluation.

\bibliographystyle{sbc}
\bibliography{references}

\end{document}